\documentclass[a4paper,11pt]{iopart}
\usepackage{iopams}
\usepackage{amssymb,amsthm}
\usepackage{amsfonts}
\usepackage{mathtools}
\usepackage{graphicx,subfigure}
\usepackage{epsfig}
\usepackage{multirow}
\usepackage[toc]{appendix}
\usepackage{cite}
\usepackage{color}
\usepackage{srcltx}
\usepackage{soul}
\usepackage{cancel}
\usepackage{ulem}

\newcommand{\A}{\mathcal{A}}
\newcommand{\C}{\mathcal{C}}
\newcommand{\Cm}{\textbf{C}}
\newcommand{\me}{\mathrm{e}}

\begin{document}
\title[Dispersion of the time spent in a state]{Dispersion of the time spent in a state: General expression for unicyclic model 
and dissipation-less precision}
\author{Somrita Ray and Andre C. Barato}
\address{Max Planck Institute for the Physics of Complex Systems, N\"othnitzer Stra\ss e 38, 01187 Dresden, Germany}

\begin{abstract}
We compare the relation between dispersion and dissipation for two random variables that 
can be used to characterize the precision of a Brownian clock. The first random variable 
is the current between states. In this case, a certain precision requires a minimal 
energetic cost determined by a known thermodynamic uncertainty relation. We introduce 
a second random variable that is a certain linear combination of two random variables, each of which 
is the time a stochastic trajectory spends in a state. Whereas the first moment of this random variable is equal to the average
probability current, its dispersion is generally different from the dispersion associated with 
the current. Remarkably, for this second random variable a certain precision can be obtained
 with an arbitrarily low energy dissipation, in contrast to the thermodynamic uncertainty 
relation for the current. As a main technical achievement, we provide an exact expression 
for the dispersion related to the time that a stochastic trajectory spends in a cluster of states
 for a general unicyclic network. 
\end{abstract}


\maketitle
\section{Introduction}

Small physical systems out of equilibrium with large fluctuations often have to be precise. 
Intuitively, for such dissipative systems the precision increases with more energy dissipation. 
The relation between fluctuations (or precision) and dissipation is an active topic of research in 
stochastic thermodynamics \cite{bara15a,bara15,piet16,piet16a,ging16,pole16,ging16a,rots16,dahr16,rold15,neri16,garr17}. There are several 
examples from biophysics \cite{qian07,lan12,meht12,bara13b,palo13,gove14,gove14a,hart15,sart14,bara14a,bo15,ito15,mcgr17} 
of the relation between energy dissipation and how ``accurate'' the particular system is, with the ``accuracy'' characterized by different mathematical objects.

A natural thought experiment in such context is to imagine  a Brownian clock \cite{bara16}, which has a pointer that on average moves in the clockwise direction, 
but for a given trajectory can move in the opposite direction due to thermal fluctuations. To give the clock a clear
physical interpretation the different positions of the pointer can be seen as different states of an enzyme 
that is driven to perform cycles in the clockwise direction by the consumption of ATP. Given modern experiments 
with single molecules, colloidal particles, and small electronic systems, such thought experiment could be realized in the 
laboratory. Furthermore, our Brownian clock could be an enzyme that controls a biochemical oscillation, which are of 
central importance for living systems \cite{ferr11,cao15,dong08}.

If our clock is modeled as a Markov process on a ring, i.e., a unicyclic network of states, we can characterize its precision by calculating the 
dispersion associated with a current random variable. This random variable is a standard choice for quantifying 
the precision of the clock, since, {\sl inter alia}, it is related to the entropy production of stochastic thermodynamics \cite{seif12}, which quantifies 
the energy dissipation of the clock. The calculation of the dispersion of this random variable for a unicyclic network in terms of the transition rates 
has been done by Derrida \cite{derr83} (see also \cite{koza99}).

In principle, the precision of the clock can also be quantified by a different random variable, like the time the stochastic trajectory spends in 
a state \cite{bara15b}. For instance this random variable is analyzed in the problem of a cell that estimates an external ligand concentration 
\cite{bara15b,berg77,bial05,endr08,endr09,mora10,gove12,kaiz14}. For the case of a unicyclic network, this random variable has been analyzed  
for a biochemical timer \cite{li02} and for a dissipative receptor estimating an external ligand concentration \cite{lang14}.

As our main technical result, we obtain an expression for the dispersion of this random variable in terms of the transition rates for a unicyclic network with arbitrary 
dynamics that does not necessarily fulfill detailed balance. We propose the use of a particular linear combination of the time a stochastic trajectory spends in a state 
to characterize the precision of a Brownian clock. For the current random variable, the thermodynamic uncertainty relation from \cite{bara15a} establishes that 
a certain precision of the clock requires a minimal amount of energy dissipation. We show that for this time random variable, the energetic cost of a certain 
precision can be arbitrarily low. Our result demonstrates that the tradeoff between precision of a Brownian clock and energy dissipation can be fundamentally different 
depending on the random variable that we choose to quantify the precision of the clock.

The paper is organized as follows. In Sec. \ref{sec2} we define the model and random variable analyzed in the paper. 
Sec. \ref{sec3} contains an expression in terms of the transition rates for the dispersion of the time random variable for 
a unicyclic network. We demonstrate that the precision quantified by the time random variable can have an arbitrarily 
low energetic cost in Sec. \ref{sec4}. We conclude in Sec. \ref{sec5}. \ref{app} contains the calculations that lead
to the expressions in Sec. \ref{sec3}.

\section{Model  definition and random variable} 
\label{sec2}

We consider a continuous time Markov process with a finite number of states $N$ in a unicyclic network. The transition rate from state $m$ to $m+1$ is denoted $w_{m}^{+}$ and the transition 
rate from state $m$ to $m-1$ is denoted $w_{m-1}^{-}$. This Markov process is represented by the scheme
\begin{equation}
1\xrightleftharpoons[w_1^-]{\,w_1^+\,}2\xrightleftharpoons[w_2^-]{\,w_2^+\,}3\; \cdots \;N-1\xrightleftharpoons[w_{N-1}^-]{\,w_{N-1}^+\,}N\xrightleftharpoons[w_{N}^-]{\,w_{N}^+\,}1.
\label{eq:scheme}
\end{equation}
The time evolution of the probability $P_m(t)$ of being in state $m$ at time $t$ for such a unicyclic scheme can be determined by the master equation
\begin{equation}
\frac{d}{d t}\textbf{P}(t)=\mathcal{L}\;\textbf{P}(t),
\label{eq:meqm}
\end{equation}
\noindent
where $\textbf{P}(t)$ is the $N$-dimensional occupation probability vector with components $P_m(t)$ and $\mathcal{L}$ is the stochastic matrix given by
\begin{equation}
\left[\mathcal{L}\right]_{ij}=
\begin{cases}
-\left(w_i^++w_{i-1}^-\right)& \text{if}\ j=i \\
w_i^+ & \text{if}\ j=i+1 \\
w_{i-1}^- & \text{if}\ j=i-1 \\
0 & \text{otherwise},
\end{cases}
\label{eq:lz0}
\end{equation}
\noindent
where $i+1=1$ for $i=N$ and $i-1=N$ for $i=1$. 

Given a stochastic trajectory from time $0$ to time $t$, the random variable $\tau_m$, which is often referred to as empirical density \cite{bara15c}, is the time 
the stochastic trajectory spends in state $m$. If we denote the state of the system at time $t'$ by $X_{t'}$, this random variable can be defined as 
\begin{equation}
\tau_m\equiv\int_0^{t}\delta_{X_{t'},m}dt'.
\end{equation}
A general linear combination of such random variable
is defined as 
\begin{equation}
\tau=\sum_{m=1}^N \alpha_m\tau_m,
\label{eq:taud}
\end{equation}
where $\alpha_m$ is an arbitrary constant. For a long time interval $t$, much larger than the time to relax to a stationary state, the first moment associated with $\tau$ is given by
\begin{equation}
v_\tau\equiv \frac{\langle\tau\rangle}{t}= \sum_{m=1}^N \alpha_m P_m,
\label{eq:jdef}
\end{equation}
where the brackets mean an average over stochastic trajectories and $P_m$ is the stationary probability to be in state $m$.

The dispersion $D_\tau$ associated with the random variable $\tau$ is defined as
\begin{equation}
D_\tau\equiv \frac{\left<\tau^2\right>-\left<\tau\right>^2}{2t}.
\label{eq:ddef}
\end{equation}
In the next section we show an exact expression for $D_\tau$ in terms of the transition rates. The derivation of this expression can be found in \ref{app}.

\section{General expression for the dispersion}
\label{sec3}

First we define the escape rate from state $m$  
\begin{equation}
\lambda_m\equiv(w_m^++w_{m-1}^-)
\label{eq:notn2}
\end{equation}
and the product of the transition rates between states $m$ and $m+1$
\begin{equation}
\nu_m\equiv w_m^+w_m^-\;.
\label{eq:notn1}
\end{equation}
The expression for the dispersion of the random variable $\tau$ defined in Eq. \eqref{eq:taud} in terms of the transition rates requires several indices, summations, and a matrix $\textbf{M}$.
We introduce these quantities below before showing this expression.

The main set of indices is $(j,n,k)$. The index $j$ is either $N$ or $N-2$, $n$ can take the values $0,1,2$, and $k$ can take the values $(N-1), (N-2), (N-3),(N-4)$. These indices 
are subjected to the constraints $n\leq j$ and  $k\leq (j-n)$. The indices $k_1$ and $k_2$ are defined by the relation $k=k_1+2 k_2$. For a given $k$ there are $\lfloor \frac{k}{2} \rfloor+1$ 
different set of values of ($k_1,k_2$), from $k_2=0$ up to $k_2=\lfloor \frac{k}{2} \rfloor$, where $\lfloor \frac{k}{2} \rfloor$ is the integer part of $k/2$. The vector $\bf{L}$ has
$k_2$ components, $\textbf{L}\equiv\{l_1,l_2,....,l_{k_2}\}$, where these indices follow the constraint $l_1\ge 1$ and $l_i+2\le l_{i+1}$, for $i=1,2,\ldots,k_2-1$.

The sets $\Omega_j\equiv\{1,2,...,j\}$ and $\Omega(\textbf{L})\equiv\{l_1,l_1+1,l_2,l_2+1,...,l_{k_2},l_{k_2}+1\}$ lead to the key set with $j-2 k_2$ integers 
\begin{equation}
\bar{\Omega}(j,\textbf{L})\equiv\Omega_j-\Omega(\textbf{L}).
\label{eq:set1}
\end{equation}
The matrix $\textbf{M}_{\Omega,i}$ for a generic set $\Omega$ with $h$ elements and the natural number $i\leq h$ is constructed in the following way. The rows of the matrix are all possible 
combinations of $i$ integers out of the set $\Omega$, where there are a total of $\Cm(h,i)\equiv\frac{h!}{(h-i)!i!}$ such combinations. The rows of the matrix are enumerated in an increasing 
order of a natural number that has $i$ digits determined by the elements of $\Omega$. Hence, the first row corresponds to the combination with the smallest such number with $i$ digits. 
Furthermore, the elements of each combination are enumerated in an increasing order. 

For example, for $(j,n,k)=(6,1,4)$ and $k_2=1$, $\bf{L}$ has only one component 
$l_1=1,2,\dots ,5$. Setting $l_1=1$ we obtain $\Omega(\textbf{L})=\{1,2\}$, which leads to $\bar{\Omega}(j,\textbf{L})=\{3,4,5,6\}$. 
For this case, $k_1=k-2 k_2=2$ and the matrix $\textbf{M}_{\bar{\Omega}(j,\textbf{L}),k_1}$ is given by
\begin{equation}
\textbf{M}_{\bar{\Omega}(j,\textbf{L}),k_1}=
\quad
\begin{pmatrix} 
3 & 4 \\
3 & 5 \\
3 & 6 \\
4 & 5 \\
4 & 6 \\
5 & 6
\end{pmatrix}.
\label{eq:m_matrix}
\end{equation}
\noindent

We introduce the following subset of $\bar{\Omega}(j,\textbf{L})$, 
\begin{equation}
\bar{\Omega}_2(j,\textbf{L},p_\lambda)\equiv\bar{\Omega}(j,\textbf{L})
-\{[\textbf{M}_{\bar{\Omega}(j,\textbf{L}),k_1}]_{p_\lambda,1},[\textbf{M}_{\bar{\Omega}(j,\textbf{L}),k_1}]_{p_\lambda,2},\dots,[\textbf{M}_{\bar{\Omega}(j,\textbf{L}),k_1}]_{p_\lambda,k_1}\},
\label{eq:set2}
\end{equation}
where $p_\lambda=1,\ldots,\Cm(j-2k_2,k_1)$. For this set we consider the matrix $\textbf{M}_{\bar{\Omega}_2(j,\textbf{L},p_\lambda),n}$.
For example, for the set $\bar{\Omega}(j,\textbf{L})$ and index $k_1$ associated with Eq. (\ref{eq:m_matrix}), for $p_{\lambda}=1$, we have 
\begin{equation}
\textbf{M}_{\bar{\Omega}_2(j,\textbf{L},p_\lambda),n}=
\quad
\begin{pmatrix} 
5 \\
6
\end{pmatrix}.
\label{eq:m2_matrix}
\end{equation}

Using the set $\bar{\Omega}(j,\textbf{L})$ in Eq. \eqref{eq:set1}, the set $\bar{\Omega}_2(j,\textbf{L},p_\lambda)$ in Eq. \eqref{eq:set2} and the matrix $\textbf{M}$ we define the sums 
\begin{align}
T^{(\alpha)}_{j,n,k,k_2}\equiv\sum_{p_\mu=1}^{\Cm(j-k,n)} \sum_{q_\mu=1}^{n}\alpha_{[\textbf{M}_{\bar{\Omega}_2(j,\textbf{L},p_\lambda),n}]_{p_\mu,q_\mu}},
\label{eq:T_alpha}
\end{align}
and 
\begin{align}
S_{j,n,k,k_2}\equiv\sum_{l_1=1}^{j-2k_2+1}\;\;\sum_{l_2=l_1+2}^{j-2k_2+3}\;\;\sum_{l_3=l_2+2}^{j-2k_2+5}...\sum_{l_{k_2}=l_{k_2-1}+2}^{j-1}\nu_{l_1} \nu_{l_2} \nu_{l_3}...\nu_{l_{k_2}}\left\{
 \sum_{p_\lambda=1}^{\Cm(j-2 k_2,k_1)} \;\;\prod_{q_\lambda=1}^{k_1} \lambda_{[\textbf{M}_{\bar{\Omega}(j,\textbf{L}),k_1}]_{p_\lambda,q_\lambda}}\right\},
\label{eq:s0def}
\end{align}
where elements of the matrix $\textbf{M}_{\bar{\Omega}(j,\textbf{L}),k_1}$, which are denoted $[\textbf{M}_{\bar{\Omega}(j,\textbf{L}),k_1}]_{p_\lambda,q_\lambda}$, 
appear in the subscript of $\lambda$ defined in Eq. \eqref{eq:notn2}, and the elements of the matrix $\textbf{M}_{\bar{\Omega}_2(j,\textbf{L},p_\lambda),n}$ appear in 
the subscript of $\alpha$ defined in Eq. \eqref{eq:taud}. Finally, the terms $b_{jn}^{(k)}$, $b^{\prime(k)}_{jn}$, and $b^{\prime\prime(k)}_{jn}$ are 
\begin{align}
&b_{jn}^{(k)}\equiv(-1)^{j-n-k}\sum_{k_2=0}^{\lfloor \frac{k}{2} \rfloor}(-1)^{k_2}S_{j,n,k,k_2}\nonumber \\
&b^{\prime(k)}_{jn}\equiv(-1)^{j-n-k}\left[\sum_{k_2=0}^{\lfloor \frac{k}{2} \rfloor}(-1)^{k_2}\left\{\theta_j-T^{(\alpha)}_{j,n,k,k_2}\right\}S_{j,n,k,k_2}\right]\nonumber\\
&b^{\prime\prime(k)}_{jn}\equiv(-1)^{j-n-k}\left[\sum_{k_2=0}^{\lfloor \frac{k}{2} \rfloor}(-1)^{k_2}\left\{\theta_j^2-2\theta_j T^{(\alpha)}_{j,n,k,k_2}
+\left[T^{(\alpha)}_{j,n,k,k_2}\right]^2\right\}S_{j,n,k,k_2}\right],
\label{eq:bjnk0}
\end{align}
\noindent
where $\theta_j\equiv\left(\alpha_1+\alpha_2+..+\alpha_j\right)$ and the primes in $b$ are related to derivatives explained in \ref{app}.

We introduce sets and summations similar to the equations above with a star superscript.  The vector $\textbf{L}^{\star}\equiv\{l^{\star}_1,l^{\star}_2,...,l^{\star}_{k_2}\}$ has components $l^{\star}_{1}\ge 2$ 
and $l^{\star}_{i}+2\le l^{\star}_{i+1}$.
Defining the sets $\Omega^{\star}_j\equiv\{2,3,...,j+1\}$ and $\Omega(\textbf{L}^{\star})\equiv \{l_1^{\star},l_1^{\star}+1,l_2^{\star},l_2^{\star}+1,...,l_{k_2}^{\star},l_{k_2}^{\star}+1\}$ we obtain
the set 
\begin{equation}
\bar{\Omega}^{\star}(j,\textbf{L})\equiv\Omega^{\star}_j-\Omega(\textbf{L}^{\star}),
\end{equation}
and the subset 
\begin{equation}
\bar{\Omega}^{\star}_2(j,\textbf{L},p_\lambda)\equiv\bar{\Omega}^{\star}(j,\textbf{L})-\{[\textbf{M}_{\bar{\Omega}^{\star}(j,\textbf{L}),k_1}]_{p_\lambda,1},
[\textbf{M}_{\bar{\Omega}^{\star}(j,\textbf{L}),k_1}]_{p_\lambda,2},\dots,[\textbf{M}_{\bar{\Omega}^{\star}(j,\textbf{L}),k_1}]_{p_\lambda,k_1}\}. 
\end{equation}
With these sets with a star subscript we define analogous sums  
\begin{align}
T^{\star(\alpha)}_{j,n,k,k_2}\equiv\sum_{p_\mu=1}^{\Cm(j-k,n)} \sum_{q_\mu=1}^{n}\alpha_{[\textbf{M}_{\bar{\Omega}^{\star}_2(j,\textbf{L},p_\lambda),n}]_{p_\mu,q_\mu}},
\label{eq:Ts_alpha}
\end{align}
and 
\begin{align}
 S_{j,n,k,k_2}^{\star}\equiv\sum_{l^{\star}_1=2}^{j-2k_2+2}\;\;\sum_{l^{\star}_2=l^{\star}_1+2}^{j-2k_2+4}\;\;\sum_{l^{\star}_3=
l^{\star}_2+2}^{j-2k_2+6}...\sum_{l^{\star}_{k_2}=l^{\star}_{k_2-1}+2}^{j}\nu_{l^{\star}_1} \nu_{l^{\star}_2} \nu_{l^{\star}_3}
...\nu_{l^{\star}_{k_2}}\left\{\sum_{p_\lambda=1}^{\Cm(j-2 k_2,k_1)} \;\;\prod_{q_\lambda=1}^{k_1} \lambda_{[\textbf{M}_{\bar{\Omega}^{\star}(j,\textbf{L}),k_1}]_{p_\lambda,q_\lambda}}\right\}.
\label{eq:ss0def}
\end{align}
\noindent
Furthermore, we introduce the respective $b^{\star}$ terms
\begin{align}
&b_{jn}^{\star (k)}\equiv(-1)^{j-n-k}\sum_{k_2=0}^{\lfloor \frac{k}{2} \rfloor}(-1)^{k_2}S_{j,n,k,k_2}^{\star}\nonumber \\
&b^{\star\prime(k)}_{jn}\equiv(-1)^{j-n-k}\left[\sum_{k_2=0}^{\lfloor \frac{k}{2} \rfloor}(-1)^{k_2}\left\{\theta_j^{\star}-T^{\star(\alpha)}_{j,n,k,k_2}\right\}S_{j,n,k,k_2}^{\star}\right]\nonumber\\
&b^{\star\prime\prime(k)}_{jn}\equiv(-1)^{j-n-k}\left[\sum_{k_2=0}^{\lfloor \frac{k}{2} \rfloor}(-1)^{k_2}\left\{\theta_j^{\star 2}-2\theta_j^{\star} T^{\star(\alpha)}_{j,n,k,k_2}+\left[T^{\star(\alpha)}_{j,n,k,k_2}\right]^2\right\}S_{j,n,k,k_2}^{\star}\right],
\label{eq:bjnks0}
\end{align}
where $\theta_j^{\star}=\left(\alpha_2+\alpha_3+..+\alpha_{j+1}\right)$.

Finally, the expressions for $v_{\tau}$ and $D_{\tau}$ in terms of the transition rates are
\begin{equation}
v_\tau=-\frac{\tilde{c_0}^{\prime(N-1)}+\tilde{c_1}^{\prime(N-1)}}{\tilde{c_1}^{(N-1)}}
\label{eq:jex}
\end{equation}
\noindent
and
\begin{equation}
D_\tau=-\frac{1}{2}\left\{\frac{\tilde{c_0}^{\prime\prime(N-2)}+\tilde{c_1}^{\prime\prime(N-2)}+\tilde{c_2}^{\prime\prime(N-2)}
+2v_{\tau}\left(\tilde{c_1}^{\prime(N-2)}+2\tilde{c_2}^{\prime(N-2)}\right)
+2v_{\tau}^2\tilde{c_2}^{(N-2)}}{\tilde{c_1}^{(N-1)}}\right\},
\label{eq:dex2}
\end{equation}
where
\begin{align}
&\tilde{c_n}^{(k)}\equiv b_{Nn}^{(k)}-\nu_{N}b_{(N-2)n}^{\star(k-2)},\nonumber\\
&\tilde{c_n}^{\prime(k)}\equiv b_{Nn}^{\prime(k)}-\nu_{N}\left[b_{(N-2)n}^{\star\prime(k-2)}\; +(\alpha_{1}+\alpha_{N})b_{(N-2)n}^{\star(k-2)}\right],\nonumber\\
&\tilde{c_n}^{\prime\prime(k)}\equiv b_{Nn}^{\prime\prime(k)}-\nu_{N}\left[b_{(N-2)n}^{\star\prime\prime(k-2)}\;
 +2(\alpha_{1}+\alpha_{N})\;b_{(N-2)n}^{\star\prime(k-2)}+(\alpha_{1}+\alpha_{N})^2\;b_{(N-2)n}^{\star(k-2)}\right].
\label{eq:cnk}
\end{align}
This final expression for the dispersion $D_{\tau}$ in terms of the transition rates is the main technical result of this paper.

\section{Relation between precision and dissipation}
\label{sec4}

First, we discuss the thermodynamic uncertainty relation
for the current variable from \cite{bara15a}. Second, we introduce the time random variable and obtain the relation between energy dissipation and
uncertainty for this random variable.

\subsection{Thermodynamic uncertainty relation for current}

The current random variable $J$ is a functional of the stochastic trajectory defined in the following way. If the pointer of the clock makes a transition from state $1$ to state $2$, this random variable 
increases by one. If the clock makes a transition in the opposite direction, from state $2$ to state $1$, this variable decreases by one. This random variable is arguably the most natural way of counting 
time with a Brownian clock, with a marker between a pair of states that counts clockwise transitions as positive and anti-clockwise transitions as negative.

The average of the current for a clock that operates for a time $t$  is given by
\begin{equation}
v_J\equiv\frac{\langle J\rangle}{t}=P_1w_1^+-P_2w_1^-,
\label{eqvJ}
\end{equation}
where the subscript $J$ is used to differentiate with the time random variable $\tau$. This quantity gives the average velocity of the clock, i.e., $v_J^{-1}$ is the average time the clock needs to 
complete a full revolution in the clockwise direction.

The uncertainty of the clock is given by
\begin{equation}
\epsilon_J^2\equiv \frac{\langle J^2\rangle-\langle J\rangle^2}{\langle J\rangle^2}= \frac{2D_J}{v_J^2t},
\label{equncJ}
\end{equation}
where
\begin{equation}
D_J\equiv \frac{\langle J^2\rangle-\langle J\rangle^2}{2t}.
\end{equation}
Similar to the time variable $\tau$, the current variable has diffusive behavior, with an uncertainty square that 
decays as $t^{-1}$. Running the clock for a longer time $t$ leads to higher precision.

The energetic cost of the clock is characterized by the rate of entropy production $\sigma$, which for the unicyclic model reads \cite{seif12} 
\begin{equation}
\sigma= \A v_J,
\end{equation}
where 
\begin{equation}
\A\equiv \ln\frac{\prod_{m=1}^{N}w_m^+}{\prod_{m=1}^{N}w_m^-}.
\end{equation}
The affinity $\A$ is the thermodynamic force that drives the system out of equilibrium. If our clock is driven by ATP,
$\A$ is the free energy liberated by the hydrolysis of one ATP, where in this paper we set Boltzmann constant $k_B$
multiplied by the temperature $T$ to $k_BT=1$. The entropy production is then the rate at which heat is dissipated by 
the clock. The total cost of operating the clock for a time $t$ is 
\begin{equation}
\C= \sigma t.
\end{equation}
Hence, the energetic cost of running the clock increases with the time $t$.

The tradeoff between energy dissipation and precision is quantified by the time independent product \cite{bara15a} 
\begin{equation}
\C\epsilon_J^2= \frac{2 D_J\sigma}{v_J^2}\ge 2.
\label{eq:thermounc}
\end{equation}
This inequality is the thermodynamic uncertainty relation from \cite{bara15a}.
It establishes that an uncertainty $\epsilon_J$ must be accompanied by the dissipation of at least 
$2/\epsilon_J^2$. We note that in equilibrium, i.e., $\A=0$, the energetic cost is zero and the uncertainty 
$\epsilon_J$ in Eq. \eqref{equncJ} diverges due to $v_J=0$. Even though we restrict our discussion
to a unicyclic network, this uncertainty relation is valid for any Markov process with a finite number of 
states \cite{bara15a}.

\subsection{Time variable vs. current}

We now introduce another random variable that can characterize the precision of the clock. Two basic requirements that this random variable 
must fulfill are: diffusive behavior with an uncertainty square that decays as $t^{-1}$ and its average divided by $t$ must be equal to
the velocity of the clock $v_J$. A linear combination of the form given in Eq. \eqref{eq:taud} that fulfills this second requirement is 
\begin{equation}
\tau\equiv w_1^+\tau_1-w_1^-\tau_2,
\label{eq:time2}
\end{equation}
where this $\tau$ is dimensionless since the rates have dimension of $t^{-1}$. In this case, 
the average $v_{\tau}$ in Eq. (\ref{eq:jdef}) becomes $v_{\tau}=v_J$, with $v_J$ given in Eq. (\ref{eqvJ}). This random variable accounts for a
different procedure to count time with the Brownian clock. Instead of a simple marker that counts transitions between $1$ and $2$
with the appropriate sign, this counting procedure would need an observer to keep track of how long the clock spends in state $1$, how long it 
spends in state $2$, and the observer must know the value of the transition rates $w_1^+$ and $w_1^-$.

Whereas the first moment of both random variables are equal, their dispersions are in general different. If we choose the random variable $\tau$ to characterize the precision of our 
Brownian clock the uncertainty becomes
\begin{equation}
\epsilon_\tau^2\equiv \frac{\langle \tau^2\rangle-\langle \tau\rangle^2}{\langle \tau\rangle^2}= \frac{2D_\tau}{v_\tau^2t}.
\end{equation}

\begin{figure}
\subfigure[]{\includegraphics[width=90mm]{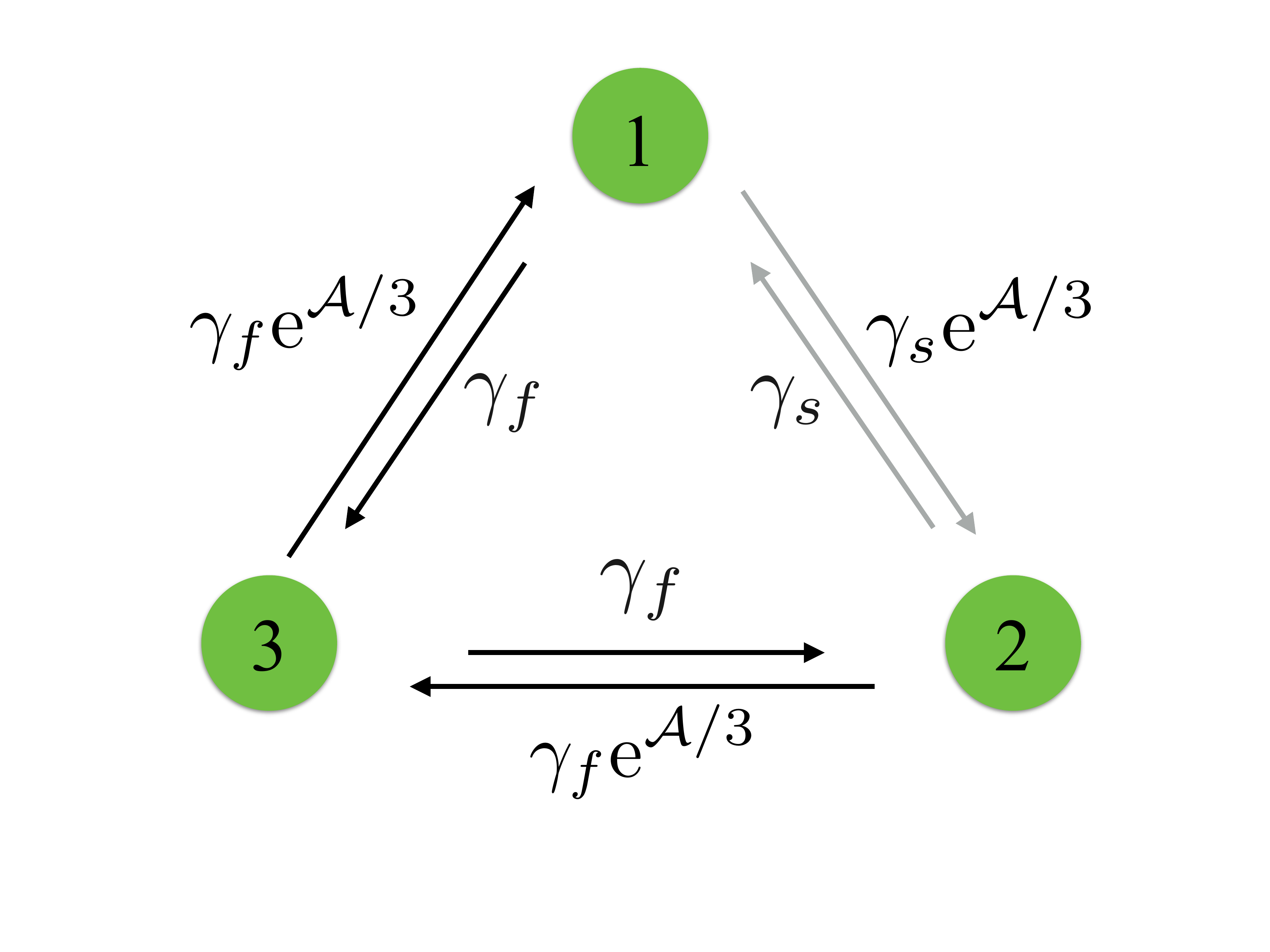}\label{fig1a}}
\subfigure[]{\includegraphics[width=75mm]{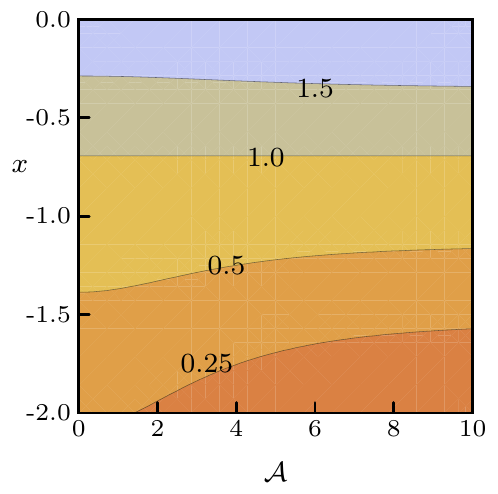}\label{fig1b}}
\vspace{-3mm}
\caption{Brownian clock with three states. (a) Particular transition rates for the calculations. (b) Contour plot of $D_\tau/D_J$ for $\gamma_f=1$, where $\gamma_s=\textrm{e}^x$.
The black lines represent the values $D_\tau/D_J= 0.25, 0.5, 1.0, 1.5$. 
}
\label{fig1}
\end{figure}

Setting $N=3$ and the transition rates to the values shown in Fig. \ref{fig1a}, where $\gamma_s$ gives the time-scale for 
transitions between states $1$ and $2$ and $\gamma_f$ gives the time scale for the other transitions, we obtain 
the following expressions for the dispersions
\begin{equation}
D_{\tau}=\frac{\gamma_s^2\left(1+\me^{\A/3}\right)
\left[\gamma_f^2\me^{\A/3}+\gamma_s\gamma_f\left(3-2\me^{\A/3}+3\me^{2\A/3}\right)+4\gamma_s^2\me^{\A/3}\right]}{\left(1+\me^{\A/3}+\me^{2\A/3}\right)\left(\gamma_f+2 \gamma_s\right)^3}
\label{eq:dtau}
\end{equation}
and
\begin{equation}
D_J=\frac{\gamma_s\gamma_f(1+\me^{\A/3})\left[\gamma_f^2(1-\me^{\A/3}+\me^{2\A/3})+4\gamma_s\gamma_f\me^{\A/3}+2\gamma_s^2(1+\me^{2\A/3})\right]}{2(1+\me^{\A/3}+\me^{2\A/3})(\gamma_f+2\gamma_s)^3},
\label{eq:dJ}
\end{equation}
where the first expression follows from Eq. (\ref{eq:dex2}) and the second expression can be calculated with the methods explained in \cite{bara15b}.

In Fig. \ref{fig1b}, we show a contour plot of the ratio $D_\tau/D_J$. If this ratio is smaller than 1 then the time random variable gives 
a higher precision than the current random variable. For small $\gamma_s$ the time random variable becomes more precise than the current 
random variable, with the crossover at $D_\tau/D_J=1$ displaying a weak dependence on the affinity $\A$.

In order to analyze the relation between dissipation and precision for this second random variable we consider the product   
\begin{equation}
\C \epsilon_\tau^2= \frac{2D_\tau\sigma}{v_\tau^2}.
\end{equation}
The average probability current in the stationary state is
\begin{equation}
v_{\tau}=\frac{\gamma_s \gamma_f\left(\me^{\A/3}-1\right)}{\gamma_f+2 \gamma_s},
\label{eq:jtau}
\end{equation}
which gives 
\begin{equation}
\C \epsilon_\tau^2=\frac{2 \A\gamma_s\left(1+\me^{\A/3}\right)\left[\gamma_f^2\me^{\A/3}+\gamma_s\gamma_f\left(3-2\me^{\A/3}+3\me^{2\A/3}\right)+4\gamma_s^2\me^{\A/3}
\right]}{\gamma_f\left(\me^{\A/3}-1\right)\left(1+\me^{\A/3}+
\me^{2\A/3}\right)\left(\gamma_f+2 \gamma_s\right)^2}.
\label{eq:qtau}
\end{equation}
For $\gamma_s\ll\gamma_f$ the product goes to zero as $\C\epsilon_\tau^2\sim \gamma_s/\gamma_f$. Hence, in the limit $\gamma_s/\gamma_f\to 0$ the energetic cost of an uncertainty $\epsilon$ 
can be arbitrarily low, in contrast to the thermodynamic uncertainty relation expressed in Eq. \eqref{eq:thermounc}. This drastic difference constitutes the main physical insight of 
this paper. We have explicitly demonstrated that the fundamental tradeoff between energy and precision depends on the random variable we select.

This fundamental difference comes from the different scaling of the dispersion $D_J$ and $D_\tau$ in the limit $\gamma_s\ll\gamma_f$. The dispersion 
$D_J$ scales in the same way as the current $v_J$, with $D_J\sim \gamma_s$. However, the dispersion $D_{\tau}$ scales as $D_\tau\sim \gamma_s^2/\gamma_f$. 
This relation is easy to understand. The dispersion associated with $\tau_m$, for $m=1,2,3$, is proportional to the fast time-scale $\gamma_f^{-1}$, 
since the escape rates of all states are of order $\gamma_f$. Since the constants in Eq. \eqref{eq:time2} are the slow rates, 
the dispersion $D_\tau$ scales with the square of the constants $\gamma_s^2$ multiplied by the fast time scale $\gamma_f^{-1}$.

An essential difference between $\tau$ and $J$ is the following. Kirchhoff's law is valid for $J$ even at the level of stochastic trajectories \cite{bara12a}. Hence, not only the average current 
is the same for the three links in the model but the dispersion is also independent of the link we choose to count the transitions. Both $v_J$ and $D_J$ are proportional to $\gamma_s$.
For the random variable $\tau$, the average velocity does not depend on the link, i.e., if we take $\tau'\equiv\tau_2w_2^+-\tau_3w_2^-$, its average is the same as the average of $\tau$ in Eq. 
\eqref{eq:time2}. However, the dispersion related to $\tau$ is generally different from the dispersion related to $\tau'$.

Even though our result was obtained with a simple three state model, it constitutes a general principle valid for any Markov process:
there is no minimal amount of energy that must be consumed in order to achieve a certain precision quantified by a random variable of the form 
given in Eq. \eqref{eq:time2}. We can simply follow the same strategy for arbitrary $N$ in a unicyclic network. Setting the rates associated 
with the link related to $\tau$ much slower than the other transition rates leads to a vanishing product $\C \epsilon_\tau^2$. 
Since a multicyclic network can always have rates in such a way that it behaves effectively like a unicyclic network \cite{bara15}, this result also 
extends to arbitrary networks.

The fact that the time random variable can lead to a limit of dissipation-less precision should not 
be confused with the results from \cite{bara16}. In this reference, a dissipation-less clock, with its
precision characterized by the current random variable, can be
obtained for systems that are driven out of equilibrium by an external periodic protocol, which 
are different from systems driven by a fixed thermodynamic force like the one considered here.

\section{Conclusion}
\label{sec5}

We have obtained an expression in terms of the transition rates for the dispersion of the time a stochastic trajectory spends in a cluster of states 
of a unicyclic network. The unicyclic network can have an arbitrary number of states with inhomogeneous transition rates. Our expression should be 
an important tool for applications in which this random variable is an observable of interest, like for a cell that estimates the concentration of 
an external ligand.

With the help of this expression we have analyzed the tradeoff between precision and dissipation in a simple three state model. We have shown
that if the precision of a Brownian clock is characterized by the time random variable from Eq. \eqref{eq:time2}, we can have a precise 
clock that dissipates an arbitrarily small amount of energy. This result is in contrast to the thermodynamic uncertainty relation for 
the current, showing that the trade-off between precision and dissipation is fundamentally different for these two random variables.

While the current random variable seems like a more natural choice to quantify the precision of a Brownian clock, with a simple 
physical interpretation like the number of products generated in a chemical reaction, the time random variable proposed here 
is, in principle, also a valid choice to quantify the precision. It would be interesting to build models that also include the thermodynamic 
cost of an internal observer that can keep track of these random variables. Intuitively, one expects that the energetic cost of the observer 
that monitors the time random variable should be higher, as keeping track of $\tau$ seems to require a more sophisticated observer. 

Investigating the relation between precision and dissipation in more elaborate models from biophysics constitutes 
a promising direction for future research \cite{cao15}. One lesson to learn from our results is that when we talk about 
fundamental limits of precision in biological systems subjected to large fluctuations, these limits are very much dependent 
on the random variable we choose to characterize the precision. \\

{\noindent \textbf{Acknowledgements}}\newline We thank \'Edgar Rold\'an  
for carefully reading the manuscript. 

\appendix
\section{Calculation of the dispersion}
\label{app}

In this appendix we obtain the expressions from Sec. \ref{sec3} that determine the dispersion $D_\tau$.
If we discretize time with a time step $\Delta$, the random variable $\tau_m$ becomes the number
of time steps the trajectory is in state $m$.
In this case, according to the Donsker-Varadhan theory the scaled cumulant generating function associated with $\tau$ in Eq. \eqref{eq:taud}
is given by the maximum eigenvalue of a modified generator \cite{touc09}. This modified generator takes the following form 
\begin{equation}
\left[\mathcal{K}(z)\right]_{ij}=
\begin{cases}
\left[1-\Delta\left(w_i^++w_{i-1}^-\right)\right]\me^{z\alpha_i}& \text{if}\ j=i \\
\Delta w_i^+\me^{z\alpha_i} & \text{if}\ j=i+1 \\
\Delta w_{i-1}^-\me^{z\alpha_i} & \text{if}\ j=i-1 \\
0 & \text{otherwise}.
\end{cases}
\label{eq:gen}
\end{equation}
\noindent
Following Ref. \cite{bara15b}, the first and second moments of $\tau$ can also be obtained in terms of the coefficients of the
characteristic polynomial of $\mathcal{K}(z)$. Defining such coefficients as
\begin{equation}
|\mathcal{R}(z)|\equiv|y\mathcal{I}-\mathcal{K}(z)| \equiv \sum_{n=0}^N c_n(z) y^n,
\label{eq:cp_coef}
\end{equation}
\noindent
we obtain the velocity $v_{\tau}$ and dispersion $D_{\tau}$ as 
\begin{equation}
v_{\tau}=-\frac{\sum_{n=0}^N c_n^{\prime}}{\sum_{n=1}^N nc_n}
\label{eq:jdef2}
\end{equation}
\noindent
and
\begin{equation}
D_{\tau}=-\frac{1}{2}\lim_{\Delta\rightarrow0}\Delta \frac{\sum_{n=0}^N c_n^{\prime\prime}+2v_{\tau}\sum_{n=1}^N nc_n^{\prime}+v_{\tau}^2\sum_{n=2}^N n(n-1)c_n}{\sum_{n=1}^N nc_n}.
\label{eq:ddef2}
\end{equation}
\noindent
The lack of explicit $z$-dependence of the coefficients denotes evaluation at $z=0$ and the primes denote derivatives with respect to $z$. Since we are interested
in the results in continuous time, the limit $\Delta \to 0$ is taken in the above equation.\\
\indent
Each coefficient $c_n(z)$ in Eq. (\ref{eq:cp_coef}) is an $(N-n)$th order polynomial in $\Delta$, \textit{i.e.}, 
\begin{equation}
c_n(z)\equiv\sum_{k=0}^{N-n}\tilde{c_n}^{(k)}(z)\;\Delta^k.
\label{eq:cn_sum}
\end{equation}
\noindent
Combining Eqs. (\ref{eq:jdef2}) and (\ref{eq:cn_sum}), we get
\begin{align}
v_{\tau}=
-\frac{\sum_{k=0}^N\left(\sum_{n=0}^{N-k}\tilde{c_n}^{\prime(k)}\right)\Delta^k}{\sum_{k=0}^N\left(\sum_{n=1}^{N-k}n\tilde{c_n}^{(k)}\right)\Delta^k}.
\label{eq:jdel}
\end{align}
\noindent
The above equation express $v_{\tau}$ as a ratio of two polynomials in $\Delta$. We find that for all $k\neq N-1$,
the coefficients of these two polynomials vanish and only the contribution of order $\mathcal{O}(\Delta^{N-1})$ survives, 
which leads to Eq. (\ref{eq:jex}). Combining Eqs. (\ref{eq:ddef2}) and (\ref{eq:cn_sum}), we obtain
\begin{align}
D_{\tau}=
-\lim_{\Delta \to 0}\Delta\left[\frac{\sum_{k=0}^N\left(\sum_{n=0}^{N-k}\tilde{c_n}^{\prime\prime(k)}+2v_{\tau}\sum_{n=1}^{N-k} n\tilde{c_n}^{\prime(k)}
+v_{\tau}^2\sum_{n=2}^{N-k} n(n-1)\tilde{c_n}^{(k)}\right)\Delta^k}
{2\sum_{k=0}^N\left(\sum_{n=1}^{N-k}n\tilde{c_n}^{(k)}\right)\Delta^k}\right],
\label{eq:jdel2}
\end{align}
\noindent
where $v_{\tau}$ is given by Eq. (\ref{eq:jex}). In this case, only the terms of order $\mathcal{O}(\Delta^{N-2})$ or higher do not vanish in the numerator, while in the denominator the sole 
contribution comes from the terms of order $\mathcal{O}(\Delta^{N-1})$, which leads to Eq. (\ref{eq:dex2}).

An expression for the coefficients $\tilde{c_n}^{(k)}(z)$ as functions of the transition rates can be obtained in the following way.
From Eqs. (\ref{eq:gen}) and (\ref{eq:cp_coef}), we get
\begin{align}
\mathcal{R}(z)
\resizebox{\linewidth}{!}{\arraycolsep=3pt%
$\equiv\left(\begin{array}{@{}*8c}
y-\me^{z \alpha_1}[1-\Delta \lambda_1] &
  -\me^{z \alpha_1}\Delta w_1^- & 0 & \cdots & \cdots & \cdots & 0 &  -\me^{z \alpha_1}\Delta w_N^+ \\
-\me^{z \alpha_2}\Delta w_1^+ & y-\me^{z \alpha_2}[1-\Delta \lambda_2] &
   -\me^{z \alpha_2}\Delta w_2^-& 0 & \cdots & \cdots & \cdots & 0 \\
0 &-\me^{z \alpha_3}\Delta w_2^+  & y-\me^{z \alpha_3}[1-\Delta \lambda_3] &-\me^{z \alpha_3}\Delta w_3^- & 0 &\cdots &\cdots &0\\
\vdots & \vdots & \vdots & \vdots & \vdots & \vdots & \vdots & \vdots \\
-\me^{z \alpha_N}\Delta w_N^- & 0 & \cdots & \cdots & \cdots & 0 & -\me^{z \alpha_N}\Delta w_{N-1}^+ &
  y-\me^{z \alpha_N}[1-\Delta \lambda_N] \\
\end{array}
\right),$}
\label{eq:cpdef}
\end{align}
\noindent
where $\lambda_m\equiv (w_m^++w_{m-1}^-)$ is the escape rate from state $m$, as defined in Eq. (\ref{eq:notn2}). 
The determinant of the $(1\times1)$ submatrix of $\mathcal{R}(z)$ starting from its upper left corner is written as
\begin{align}
d_1(z)\equiv[\mathcal{R}(z)]_{11}=y-\me^{z \alpha_1}[1-\Delta \lambda_1].
\label{eq:d1}
\end{align} 
\noindent
The determinant $d_2(z)$ of the $(2\times2)$ submatrix of $\mathcal{R}(z)$ starting from its upper left corner can be expressed in terms of $d_1(z)$ as   
\begin{align}
d_2(z)=\left[y-\me^{z \alpha_2} (1-\Delta \lambda_2)\right] d_1(z)-\me^{z (\alpha_1+\alpha_2)}\Delta^2 \nu_1.
\label{eq:d2}
\end{align} 
\noindent
In a similar way, $d_j(z)$, the determinant for the $(j\times j)$ submatrix of $\mathcal{R}(z)$ starting from its upper left corner for $3 \leq j \leq N$
can be expressed by the general recursion relation \cite{chem08}
\begin{equation}
d_j(z)=\left[y-\me^{z \alpha_j} (1-\Delta \lambda_j)\right]d_{j-1}(z)
-\me^{z (\alpha_{j-1}+\alpha_j)} \Delta^2 \nu_{j-1} d_{j-2}(z), 
\label{eq:dj_rec}
\end{equation}
\noindent
where $\nu_m\equiv w_m^+w_m^-$, as defined in Eq. (\ref{eq:notn1}). This relation is valid only for a tridiagonal matrix and, hence, $d_N(z) \neq |\mathcal{R}(z)|$.

To account for the terms in $|\mathcal{R}(z)|$ that appear due to the presence of the non-zero elements $[\mathcal{R}(z)]_{1N}$ and $[\mathcal{R}(z)]_{N1}$,
we define a second kind of determinant, starting from one element left and one below from the upper left corner of $\mathcal{R}(z)$. Such determinant
for the $(1\times1)$ submatrix of $\mathcal{R}(z)$ is
\begin{equation}
d_1^{\star}(z)\equiv[\mathcal{R}(z)]_{22}=y-\me^{z \alpha_2}[1-\Delta \lambda_2].
\label{eq:d1s}
\end{equation}
\noindent
The determinant $d_2^{\star}(z)$ of the $(2\times 2)$ submatrix can be expressed in terms of $d_1^{\star}(z)$ as
\begin{equation}
d_2^{\star}(z)=\left[y-\me^{z \alpha_3} (1-\Delta \lambda_3)\right] d_1^{\star}(z)-\me^{z(\alpha_2+\alpha_3)}\Delta^2 \nu_2.
\label{eq:d2s}
\end{equation}
\noindent
In a similar manner, $d_j^{\star}(z)$, the determinant for the $(j\times j)$ submatrix of $\mathcal{R}(z)$ starting from $[\mathcal{R}(z)]_{22}$ for $3 \leq j \leq N-1$
is given by the recursion relation
\begin{equation}
d_j^{\star}(z)=[y-\me^{z \alpha_{j+1}} (1-\Delta \lambda_{j+1})]d_{j-1}^{\star}(z)
-\me^{z(\alpha_{j}+\alpha_{j+1})} \Delta^2 \nu_{j} d_{j-2}^{\star}(z)
\label{eq:djs_rec}
\end{equation}
\noindent
We can now express $|\mathcal{R}(z)|$ in terms of these two types of determinants as \cite{chem08}
\begin{equation}
|\mathcal{R}(z)|=d_{N}(z)
-d_{N-2}^{\star}(z)\;\Delta^2 \nu_{N}\me^{z(\alpha_{1}+\alpha_{N}) }-\Delta^N (w_{1}^+w_{2}^+...w_{N}^++w_{1}^-w_{2}^-...w_{N}^-)\me^{z \theta_N},
\label{eq:detR}
\end{equation}
\noindent
where $\theta_N=\alpha_{1}+\alpha_{2}+...+\alpha_{N}$.

From Eq. (\ref{eq:cpdef}), we get that $d_j(z)$ and $d_j^{\star}(z)$ are $j-th$ order polynomials in $y$. Thus we define
\begin{equation}
d_j(z)=\sum_{n=0}^{j}a_j^{(n)}(z)y^n
\label{eq:dj_sum}
\end{equation}
\noindent
and
\begin{equation}
d_j^{\star}(z)=\sum_{n=0}^{j}a_j^{\star(n)}(z)y^n.
\label{eq:djs_sum}
\end{equation}
\noindent
From Eqs. \eqref{eq:cp_coef}, (\ref{eq:detR}), (\ref{eq:dj_sum}) and (\ref{eq:djs_sum}), we obtain
\begin{equation}
c_n(z)=a_N^{(n)}(z)-\nu_{N} \Delta^2 a_{N-2}^{\star(n)}(z) \me^{z(\alpha_{1}+\alpha_{N})}-\delta_{n,0}\left[\Delta^N (w_{1}^+w_{2}^+...w_{N}^++w_{1}^-w_{2}^-...w_{N}^-)
\me^{z \theta_N}\right].
\label{eq:cn_rec}
\end{equation}
Combining Eqs. (\ref{eq:dj_rec}) and (\ref{eq:dj_sum}), we get the recursion relation
\begin{equation}
a_j^{(n)}(z)=a_{j-1}^{(n-1)}(z)
-a_{j-1}^{(n)}(z)\; \me^{z \alpha_{j}}\left[1-\Delta \lambda_j\right]-a_{j-2}^{(n)}(z)\; \Delta^2 \nu_{j-1}\me^{z (\alpha_{j-1}+\alpha_j)},
\label{eq:aj_rec}
\end{equation}	
\noindent
where $a_j^{(n)}(z) \neq 0$ for $0\le j\le N$ and $0\le n\le j$. Utilizing Eq. (\ref{eq:dj_sum}) for $j=1$, we get the initial values for the
 recursion relation as $a_1^{(0)}(z)=-\me^{z \alpha_1} [1-\Delta \lambda_1]$ and $a_1^{(1)}(z)=1$. In addition, we define $a_0^{(0)}(z)=1$.
From Eqs. (\ref{eq:djs_rec}) and (\ref{eq:djs_sum}), we establish a similar kind of recursion relation for $a_j^{\star(n)}(z)$  
\begin{equation}
a_j^{\star(n)}(z)=a_{j-1}^{\star(n-1)}(z)
-a_{j-1}^{\star(n)}(z)\; \me^{z \alpha_{j+1}}\left[1-\Delta \lambda_{j+1}\right]-a_{j-2}^{\star(n)}(z)\; \Delta^2 \nu_{j}\me^{z (\alpha_{j}+\alpha_{j+1})},
\label{eq:ajs_rec}
\end{equation}
\noindent
where $a_j^{\star(n)}(z)\neq 0$ for $0\le j\le N-2$ and $0\le n\le j$. From Eq. (\ref{eq:djs_sum}) for $j=1$, the initial values for the
above recursion are obtained as $a_1^{\star(0)}(z)=-\me^{z \alpha_2} [1-\Delta \lambda_2]$, $a_1^{\star(1)}(z)=1$
and we define $a_0^{\star(0)}(z)=1$.

Since each coefficient $a_j^{(n)}(z)$ [or $a_j^{\star(n)}(z)$] is a $(j-n)$th order polynomial in $\Delta$, we define
\begin{equation}
a_j^{(n)}(z)=\sum_{k=0}^{j-n}b_{jn}^{(k)}(z)\;\Delta^k
\label{eq:aj_sum}
\end{equation}
\noindent
and
\begin{equation}
a_j^{\star(n)}(z)=\sum_{k=0}^{j-n}b_{jn}^{\star(k)}(z)\;\Delta^k.
\label{eq:ajs_sum}
\end{equation}
\noindent
From Eqs. (\ref{eq:cn_sum}), (\ref{eq:cn_rec}), (\ref{eq:aj_sum}) and (\ref{eq:ajs_sum}), we obtain
\begin{equation}
\tilde{c_n}^{(k)}(z)=b_{Nn}^{(k)}(z)-\nu_Nb_{(N-2)n}^{\star(k-2)}(z)\;\me^{z(\alpha_{1}+\alpha_{N})}-\delta_{n,0} \delta_{k,N}\left[ (w_{1}^+w_{2}^+...w_{N}^++w_{1}^-w_{2}^-...w_{N}^-)
\me^{z\theta_N}\right],
\label{eq:cnk_rec}
\end{equation}
\noindent
where $\delta_{n,0}$ and $\delta_{k,N}$ are Kronecker deltas. 
Eqs. (\ref{eq:aj_rec}) and (\ref{eq:aj_sum}) lead to the following recursion relation 
\begin{equation}
b_{jn}^{(k)}(z)=b_{(j-1)(n-1)}^{(k)}(z)
-\left[b_{(j-1)n}^{(k)}(z)-\lambda_{j}b_{(j-1)n}^{(k-1)}(z)\right]\me^{z \alpha_{j}}-b_{(j-2)n}^{(k-2)}(z)\; \nu_{j-1}\me^{z (\alpha_{j-1}+\alpha_j)},
\label{eq:bjn_rec}
\end{equation}
where $b_{jn}^{(k)}(z)\neq 0$ for $0\leq j\leq N$, $0\leq n\leq j$ and $0\leq k\leq j-n$. For $j=0$, the only possible values of $(j,n,k)$ are $(0,0,0)$
 and for $j=1$, there are $(1,0,0)$, $(1,0,1)$ and $(1,1,0)$. Thus the initial values for the above recursion are given by the four
 coefficients $b_{00}^{(0)}(z)=1$, $b_{10}^{(0)}(z)=-\me^{z \alpha_1}$, $b_{10}^{(1)}(z)=\lambda_1\me^{z \alpha_1}$ and $b_{11}^{(0)}(z)=1$.
\noindent
In a similar manner, the coefficients $b_{jn}^{\star(k)}(z)$ are given by the recursion relation
\begin{equation}
b_{jn}^{\star(k)}(z)=b_{(j-1)(n-1)}^{\star(k)}(z)
-\left[b_{(j-1)n}^{\star(k)}(z)-\lambda_{j+1}b_{(j-1)n}^{\star(k-1)}(z)\right]\me^{z \alpha_{j+1}}-b_{(j-2)n}^{\star(k-2)}(z)\; \nu_{j}\me^{z (\alpha_{j}+\alpha_{j+1})},
\label{eq:bjnks_rec}
\end{equation}
\noindent	
where $b_{jn}^{\star(k)}(z)\neq 0$ for $0\leq j\leq N-2$, $0\leq n\leq j$ and $0\leq k\leq j-n$. The initial values in this case are $b_{00}^{\star(0)}(z)=1$,
 $b_{10}^{\star(0)}(z)=-\me^{z \alpha_2}$, $b_{10}^{*(1)}(z)=\lambda_2\me^{z \alpha_2}$ and $b_{11}^{\star(0)}(z)=1$.

The solution of Eq. (\ref{eq:bjn_rec}) is
 \begin{equation}
 b_{jn}^{(k)}(z)=(-1)^{j-n-k}g_j(z)\sum_{k_2=0}^{\lfloor \frac{k}{2} \rfloor}(-1)^{k_2}S_{j,n,k,k_2}(z),
 \label{eq:bjnk_sol}
 \end{equation}
\noindent
where $g_j(z)\equiv\me^{z(\alpha_1+\alpha_2+...+\alpha_j)}$. The  terms $S_{j,n,k,k_2}(z)$ in the above expression are given by
\begin{equation}
S_{j,n,k,k_2}(z)\equiv T^{(\nu)}_{j,k_2}\times T^{(\lambda)}_{j,k,k_2} \times
T^{(\mu)}_{j,n,k,k_2}(z),
 \label{eq:sjnk}
 \end{equation}
\noindent
where
\begin{equation}
T^{(\nu)}_{j,k_2}\equiv\sum_{l_1=1}^{j-2k_2+1}\;\;\sum_{l_2=l_1+2}^{j-2k_2+3}\;\;\sum_{l_3=l_2+2}^{j-2k_2+5}...\sum_{l_{k_2}=l_{k_2-1}+2}^{j-1}\nu_{l_1} \nu_{l_2} \nu_{l_3}...\nu_{l_{k_2}},
 \label{eq:tnu}
\end{equation}
\begin{equation}
T^{(\lambda)}_{j,k,k_2}\equiv\sum_{p_\lambda=1}^{\Cm(j-2 k_2,k_1)} \;\;\prod_{q_\lambda=1}^{k_1} \lambda_{[\textbf{M}_{\bar{\Omega}(j,\textbf{L}),k_1}]_{p_\lambda,q_\lambda}},
 \label{eq:tlam}
 \end{equation}
 \noindent
 and
\begin{equation}
T^{(\mu)}_{j,n,k,k_2}(z)\equiv\sum_{p_\mu=1}^{\Cm(j-k,n)} \;\;\prod_{q_\mu=1}^{n} \mu_{[\textbf{M}_{\bar{\Omega}_2(j,\textbf{L},p_\lambda),n}]_{p_\mu,q_\mu}}.
\label{eq:tmu}
\end{equation} 
\noindent
Here $\mu_j(z)\equiv \textrm{e}^{-z\alpha_j}$ and the matrix $\textbf{M}$ is introduced in Sec. \ref{sec3}. The solution of Eq. (\ref{eq:bjnks_rec}) is 
 \begin{equation}
  b_{jn}^{\star(k)}(z)=(-1)^{j-n-k}g^{\star}_j(z)\sum_{k_2=0}^{\lfloor \frac{k}{2} \rfloor}(-1)^{k_2}S^{\star}_{j,n,k,k_2}(z),
 \label{eq:bjnks_sol}
 \end{equation}
\noindent
where $g^{\star}_j(z)\equiv\me^{z(\alpha_2+\alpha_3+...+\alpha_{j+1})}$,
\begin{equation}
S^{\star}_{j,n,k,k_2}(z)\equiv T^{\star(\nu)}_{j,k_2}\times T^{\star(\lambda)}_{j,k,k_2} \times
T^{\star(\mu)}_{j,n,k,k_2},
\label{eq:sjnks}
\end{equation} 
\begin{equation}
T^{\star(\nu)}_{j,k_2}\equiv\sum_{l^{\star}_1=2}^{j-2k_2+2}\;\;\sum_{l^{\star}_2=l^{\star}_1+2}^{j-2k_2+4}\;\;\sum_{l^{\star}_3=l^{\star}_2+2}^{j-2k_2+6}
\cdots\sum_{l^{\star}_{k_2}=l^{\star}_{k_2-1}+2}^{j}\nu_{l^{\star}_1} \nu_{l^{\star}_2} \nu_{l^{\star}_3}...\nu_{l^{\star}_{k_2}},
\label{eq:tnus}
\end{equation}
\begin{equation}
T^{\star(\lambda)}_{j,k,k_2}\equiv\sum_{p_\lambda=1}^{\Cm(j-2 k_2,k_1)}\;\;\prod_{q_\lambda=1}^{k_1} \lambda_{[\textbf{M}_{\bar{\Omega}^{\star}(j,\textbf{L}),k_1}]_{p_\lambda,q_\lambda}},
\label{eq:tlams}
\end{equation}
\noindent
and
\begin{equation}
T^{\star(\mu)}_{j,n,k,k_2}\equiv\sum_{p_\mu=1}^{\Cm(j-k,n)} \;\;\prod_{q_\mu=1}^{n} \mu_{[\textbf{M}_{\bar{\Omega}^{\star}_2(j,\textbf{L},p_\lambda),n}]_{p_\mu,q_\mu}}.
\label{eq:tmus}
\end{equation}  
\indent
Eq. (\ref{eq:cnk_rec}) along with Eqs. (\ref{eq:bjnk_sol} $-$ \ref{eq:tmus}) leads to an expression of the coefficients $\tilde{c}^{(k)}_n(z)$ in terms of the transition 
rates. Taking derivatives of this coefficients and setting $z=0$ we obtain the expressions from Sec. \ref{sec3}.

\end{document}